\documentstyle[12pt]{article}
\begin{document}

{\large \bf \begin{center}
Large-$N$ nonlinear $\sigma$ models on $R^2\times S^1$
\end{center}}
\vspace*{1 cm}

\begin{center} Dae Yup Song\\
Department of Physics,\footnote{Permanent Address}  \\
Sunchon National University, Sunchon 540-742, Korea\\
and\\
Physics Department,\footnote{Present Address} Brown University, Providence,
RI02912
\end{center}
\vspace*{3 cm}

The large-$N$ nonlinear $O(N)$, $CP^{N-1}$ $\sigma$ models are studied on $R^2
\times S^1$.  The $N$-components scalar fields of the models are supposed to
acquire a phase $e^{i2\pi\delta}$ $(0\leq \delta <1)$, along the circulation of
the circle, $S^1$.  We evaluate the effective potentials  to the leading
order of the $1/N$ expansion. It is shown that, on $R^2\times S^1$ the $O(N)$
model has rich phase structure while the phase of $CP^{N-1}$ model is just
that of the model at finite temperature.

\newpage

\begin{center} {\bf I. INTRODUCTION}\end{center}

On 3-dimensional Euclidean spacetime, a wide class of quantum field theories
which are not renormalizable in a weak-coupling expansion are renormalizable
in the $1/N$ expansion [1-3]. These include the nonlinear $O(N)$, $CP^{N-1}$
$\sigma$ models [1,2] which are of continuous interests since they describe
two-dimensional antiferromagnets [4]. In the nonlinear $\sigma$ models there
are critical coupling constants, $g_c$, which separate the ordered and
disordered
phase.

In this paper we will study the $O(N)$ model and the $CP^{N-1}$ model on
$R^2\times S^1$, where the circumference of the circle $S^1$ is $L$. We
evaluate the effective potentials, $V_L$, to the leading
order of $1/N$ expansion to find the phase structures of the models. The
nonlinear $\sigma$ models share many properties with each other. And the phase
structures of them are identical on $R^3$ [1,2] and at finite temperature [5].

For field theories on a spacetime of nontrivial topology the boundary
condition of the fields should be specified. In this respect, we take the same
point of view of Ref.[6] ( see also our recent paper[7] ): A field could take a
phase
$e^{i2\pi\delta}$ $(0\leq \delta <1)$, along the circulation of the circle,
$S^1$. In the $CP^{N-1}$ model, there is a Abelian gauge symmetry [8] while
there
is no such symmetry in the $O(N)$ model. One of the interesting properties of
Abelian gauge theory coupled to matter field $R^n\times S^1$ is that a change
of
boundary condition of matter field is traded for a change of gauge field [6].
In the $CP^{N-1}$ model, the same phenomenon occurs and it leads to that the
model be always in the disordered phase as in the model at finite temperature
[5].

In the nonlinear $O(N)$ $\sigma$ model, however, the phase structure
depends on the $\delta$ and the coupling constant $g$; For some cases there is
critical circumference, $L_c$, which depends on $g$ and $\delta$ and separates
the ordered and disordered phase, as will be found analytically. The
$\delta\rightarrow 0$ limit is of
particular interest, since $V_L(\delta=0)$ is just the effective potential of
the model at a
temperature $T$ $(=1/k_BL)$.  For $g<g_c$, $L_c$ approaches to $\infty$ in this
limit, which corresponds to the phase transition at $T_c\rightarrow 0^+$ found
in the finite-temperature analysis of the model [9].

In the next section we will analyze the $O(N)$ model and a similar analysis
will be carried on $CP^{N-1}$ model in Sec. III. The final section is devoted
to
the discussions.

\begin{center} {\bf II. THE O(N) MODEL}\end{center}

\noindent
{\bf A. Effective potential formalism for the model on $R^3$}\newline

The Lagrangian density of the $O(N)$ model is
\begin{equation}
{\cal L}=\frac{1}{2}\partial_\mu n\partial_\mu n
            +\frac{\sigma}{2}(n^2-N/g_0^2)
\end{equation}
where $n$ is the $N$-component scalar field and $\sigma$ is a Lagrangian
multiplier for the constraint
\begin{equation}
n^2=\frac{N}{g_0^2}.
\end{equation}
The effective potential  in the leading order of the $1/N$
expansion is given for constant $\sigma$ by the tree and one-loop diagrams
with external $\sigma$
lines [10,11]  and on $R^3$ it is written as
\begin{equation}
\frac{V_0}{N}=-\frac{\sigma}{2g_0^2}+\frac{1}{2}\int_{\mid p_E\mid <\Lambda}
                 \frac {d^3 p_E}{(2\pi)^3}\ln[1+\frac{\sigma}{p_E^2}],
\end{equation}
where $\Lambda$ is the cutoff. The renormalization of $V_0$ can be done by
demanding that
\begin{equation}
\frac{1}{N}\frac{\partial V_0}{\partial\sigma}\mid_{\sigma=M^2}
 =-\frac{1}{g^2}.
\end{equation}
Then the renormalized effective potential is
\begin{eqnarray}
\frac{V_0}{N}&=&-\frac{\sigma}{2g^2}
  -\frac{\sigma}{2}\int_{\mid
p_E\mid<\Lambda}\frac{d^3p_E}{(2\pi)^3}\frac{1}{p_E^2+M^2}
   +\frac{1}{2}\int_{\mid p_E\mid <\Lambda}\frac{d^3p_E}{(2\pi)^3}
                    \ln[1+\frac{\sigma}{p_E^2}]\\
&=&\frac{\sigma}{2}[\frac{M}{4\pi}-\frac{1}{g^2}]-\frac{\sigma^{3/2}}{12\pi}
       +O(1/\Lambda).
\end{eqnarray}
For large $\sigma$, the behavior of $V_0$ is determined by the second term of
the right-hand side of Eq.(6) which decreases as $\sigma$ increases. The
necessary and sufficient condition for the existence of stationary point in
$V_0$ is that $\frac{M}{8\pi}-\frac{1}{2g^2}$ ( the first derivative of $V_0$
at $\sigma=0$ ) is positive. This condition is satisfied for the coupling
constant $g>g_c$ where the critical coupling constant $g_c$ is given as
\begin{equation}
g_c^2=\frac{4\pi}{M}.
\end{equation}

For the $g>g_c$, there is a global stationary point of $V_0$ at
\begin{equation}
\sigma=m_0^2=(M-\frac{4\pi}{g^2})^2.
\end{equation}
The presence of stationary point in effective potential which is absent for
tree approximation denotes the dynamical mass generation [10]. $m_0$ is the
dynamically generated mass of $z$-particles. On the other hand, for $g<g_c$,
there is no stationary point, which means that dynamical generation of mass
does {\em not} occur in this phase.

The saddle point analysis through effective action has been carried out
in Ref.[1] and the physical pictures of the model there agree with those
in this paper. Though we use the manifestly $O(N)$ invariant effective
potential formalism, the results of Ref.[1] imply that for $g<g_c$ ( i.e.
for the phase of no dynamical mass generation  ) the stationary condition
can be satisfied with nonvanishing $<n>$ while for $g>g_c$ ( i.e. for the phase
of dynamical mass generation ) $<n>=0$; That is, the model would be in
ordered phase for the case of no dynamical mass generation and the model is in
disordered phase when dynamical mass generation takes place. This will be
accepted [4] in this paper afterwards.
\newline

\noindent
{\bf B. Phase structure on $R^2\times S^1$}\newline

Now we will evaluate the effective potential on $R^2\times S^1$ to find the
phase structure. As in Ref.[6,7], it will be assumed that the $n$ fields are
{\em quasi}periodic under the circulation of the circle $S^1$ whose
circumference is $L$:
\begin{equation}
n(x_1, x_2, x_3+L)= e^{i2\pi\delta} n(x_1, x_2, x_3)~~~(0\leq \delta <1).
\end{equation}
In such boundary condition, the third component of the momentum of
$z$-fields has the discrete values
\begin{equation}
p_3=\frac{2\pi}{L}(n+\delta)~~~(n=0,\pm 1,\pm 2,\ldots),
\end{equation}
and the effective potential for $\sigma$ in the leading order of the $1/N$
expansion is written as
\begin{equation}
V_L(\sigma,\delta)=-\frac{\sigma}{2g_0^2}+\frac{1}{2L}\sum_{n=-\infty}^\infty
\int\frac{d^2p_E}{(2\pi)^2}\ln[1+\frac{\sigma}
      {(\frac{2\pi}{L})^2(n+\delta)^2 +p_E^2}].
\end{equation}

Making use of the formula [6,7]
\begin{eqnarray}
\sum_{n=-\infty}^\infty \ln [1+\frac{b^2}{(n+\delta)^2+a^2}]
&=&\int_{-\infty}^\infty \ln[1+\frac{b^2}{\tau^2+a^2}]d\tau\nonumber\\
&&+\ln\frac{1-2\cos(2\pi\delta)e^{-2\pi\sqrt{a^2+b^2}}+e^{-4\pi\sqrt{a^2+b^2}}}
         {1-2\cos(2\pi\delta)e^{-2\pi\mid a\mid}+e^{-4\pi\mid a\mid}},
\nonumber\\
\end{eqnarray}
the topology effect in the effective potential can be separated:
\begin{equation}
V_L(\sigma,\delta)=V_0(\sigma)+NV_\Delta(\sigma,L,\delta),
\end{equation}
where
\begin{eqnarray}
V_\Delta &=& \frac{1}{4\pi L}\int_0^\infty dp~p
\ln \frac{1-2\cos(2\pi\delta)e^{-L\sqrt{p^2+\sigma}}+e^{-2L\sqrt{p^2+\sigma}}}
    {1-2\cos(2\pi\delta)e^{-Lp}+e^{-2Lp}}\nonumber\\
   &=&\frac{1}{2\pi L^3}\sum_{n=1}^\infty\frac{\cos(2\pi\delta n)}{n^3}
         (1-e^{-nL\sqrt{\sigma}})\nonumber\\
   & &-\frac{1}{2\pi L^2}\sum_{n=1}^\infty \frac{\sqrt{\sigma}}{n^2}
          \cos(2\pi\delta n) e^{-nL\sqrt{\sigma}}.
\end{eqnarray}
$V_\Delta$, the effect of topology, is finite for any $L,\delta,$ and
$\sigma$, and reduced to zero as $L$ approach to $\infty$. Since $V_\Delta$
is finite, the renormalization of $V_L$  can be done by that of $V_0$.

If $V_L$ has stationary point at $\sigma=m_L^2$, then $m_L$ satisfies the
following equation
\begin{eqnarray}
\frac{1}{N}\frac{\partial V_L}{\partial \sigma}\mid_{\sigma=m_L^2} &=&
      0\\
&=&-\frac{1}{2g^2}+\frac{1}{8\pi}[M-m_L]+\frac{1}{4\pi L}
      \sum_{n=1}^\infty\frac{\cos(2\pi\delta n)}{n}e^{-nLm_L}\nonumber\\
&=&-\frac{1}{2g^2}+\frac{1}{8\pi}[M-m_L]\nonumber\\
& &-\frac{1}{8\pi L}\ln[1-2\cos(2\pi\delta)e^{-Lm_L}+e^{-2Lm_L}].\nonumber
\end{eqnarray}
The (formal) solution of Eq.(15) is that
\begin{equation}
m_L=\frac{1}{L}\cosh^{-1}[\frac{\exp\{(M-\frac{4\pi}{g^2})L\}
         +2\cos(2\pi\delta)}{2}].
\end{equation}
It is instructive to expand $V_L$ in terms of small $\sigma$:
\begin{equation}
\frac{V_L}{N}=\{\frac{M}{4\pi}-\frac{1}{g^2}
-\frac{1}{4\pi L}\ln(2-2\cos(2\pi\delta)\}
      \sigma +O(\sigma^{3/2})~~~~~~~(\delta\neq 0).
\end{equation}
Since $V_\Delta$ approaches to a fixed value as $\sigma$ goes to
$\infty$, for large $\sigma$ the behavior of $V_L$  is determined by that
of $V_0$ which decrease as $\sigma$ increase. A sufficient condition for
the existence of stationary point  is, therefore, that $V_L$ increases
as $\sigma$  increases in the vicinity of $\sigma=0$. The
condition can be obtained from Eq.(17) as:
\begin{equation}
\frac{M}{4\pi}-\frac{1}{g^2}-\frac{1}{4\pi L}\ln(2-2\cos(2\pi\delta))~ >~0,
\end{equation}
which is just the condition for the existence of $m_L$ in Eq.(16).

In the $\delta\rightarrow 0$ limit, the condition (18) is satisfied for any
$g$ and finite $L$, which agrees with the result that the model is in
disordered phase at any finite temperature [9]. For $\delta=0$ it is easy to
find a
formula familiar through the finite-temperature analyses ( for example, see
Ref.[5] ),
\[
\frac{1}{N}\frac{\partial V_L(\delta=0)}{\partial \sigma}
= -\frac{1}{2g^2}+\frac{1}{8\pi}[M-\sqrt{\sigma}]
-\frac{1}{4\pi L} \ln[1-e^{-L\sqrt{\sigma}}],
\]
where the $V_L(\delta=0)$ is the effective potential of the model at a
temperature $T=1/k_B L$.

To discuss the phase structure, it is convenient to consider the cases
$g>g_c~(~{\rm that~ is, }~ \frac{M}{4\pi}-\frac{1}{g^2}>0)$, $g<g_c$,
separately.

\begin{center}{\bf a. $g>g_c$}\end{center}

On $R^3$, for this coupling constant the $z$-fields have the dynamically
generated mass $m_0$ of Eq.(8) and the model is in disordered phase. On
$R^2\times S^1$, there are two cases:\newline
(i) For $\frac{1}{6}<\delta<\frac{5}{6}$, there is a critical circumference
of the $S^1$, $L_c$, which is given as
\begin{equation}
L_c=\frac{1}{M-\frac{4\pi}{g^2}}\ln\{2-2\cos(2\pi\delta)\}.
\end{equation}
If $L$ is smaller than $L_c$, there is no dynamical mass generation and the
model is in ordered phase, which is absent for $g>g_c$ on $R^3$. If $L$ is
larger than $L_c$, the dynamical mass generation takes place and the model is
in disordered phase as in the model on $R^3$. $m_L$ in Eq.(16) is the mass of
$z$-particles. \newline
(ii) For $\delta$ not mentioned in (i), the dynamical mass generation occurs
for all $L$ and the model is in disordered phase as in the model on $R^3$.
Again, the mass of $z$-particles is given as $m_L$ in Eq.(16).

\begin{center}{\bf b. $g<g_c$}\end{center}

\noindent
(i) For $\frac{1}{6}<\delta<\frac{5}{6}$, there is no dynamical mass
generation for all $L$ and the model is in ordered phase, as in the model on
$R^3$.\newline
(ii) For $\delta$ not mentioned in (i), the critical  circumference  $L_c$
in Eq.(19) is positive and separate the ordered and disordered phases.
For $L<L_c$ the model is in disordered phase and the
mass of $z$-particle is given as $m_L$  in Eq.(16), which is absent for the
model of
$g<g_c$ on $R^3$. When $L>L_c$, there is no dynamical mass generation.
In the $\delta\rightarrow 0$ limit, $L_c$ approaches to $\infty$ which denote
the phase
transition  at $T_c\rightarrow 0^+$ of Ref.[9].

In every case of {\bf a.} and {\bf b.}, when $L$ approaches to $\infty$
the results of the model on $R^2\times S^1$ reproduce those on $R^3$.

\begin{center}{\bf III. THE $CP^{N-1}$ MODEL}\end{center}

In this section, to denote this model we use the same notations used for the
$O(N)$ model. This will go on throughout the paper unless there
is confusion.\newline

\noindent
{\bf A. Effective potential formalism of the model on $R^3$}\newline

The model is described by the Lagrangian density
\begin{equation}
{\cal L}=(\partial_\mu-iA_\mu)z^\dagger(\partial_\mu + iA_\mu)z
         +\sigma(z^\dagger z-N/g_0^2),
\end{equation}
where $z$ is an $N$-component complex scalar field. At the classical level,
as in the 2-dimensional model [8],  $A_\mu$ is an auxiliary field which can
be replaced by
\[ \frac{ig_0^2}{2N}[z^\dagger\partial_\mu z-(\partial_\mu^\dagger)z], \]
and $\sigma$ is a Lagrangian multiplier for the constraint
\begin{equation}
z^\dagger z=\frac{N}{g_0^2}.
\end{equation}

The Abelian gauge symmetry of the model is that the substitutions of $z$ and
$A_\mu$ fields by $e^{i\alpha(x)}z$ and $A_\mu-\partial_\mu\alpha(x)$
respectively do not cause change for the Lagrangian density of Eq.(20).
Because of the gauge fields, we use a different method for evaluating
effective potential. The Lagrangian density can be written as
\[ {\cal L}=z^\dagger D z -N\sigma/g_0^2,  \]
where
\[ D=-(\partial_\mu+iA_\mu)(\partial_\mu + iA_\mu)+\sigma. \]
The Gaussian functional integral for the effective potential gives the
radiative contribution, $({\rm Tr}~\ln D+ C)$ divided by the spacetime
volume [12], or
\begin{equation}
\frac{V_0}{N}=-\frac{\sigma}{g_0^2}+{\rm Tr}~\ln D + C.
\end{equation}
$C$  is constant which may arise from functional integral and it will be fixed
by
demanding that $V_0\mid_{\sigma=0}=0$. In $1/N$ expansion, the diagrams which
have $\sigma$ or $A_\mu$ propagator as internal lines give contribution of
next to the leading order and can be ignored in the large-$N$ analysis. This
means that the $\sigma$ and $A_\mu$ look like spacetime constants in the
large-$N$ approximation. These considerations give the effective potential
\begin{equation}
\frac{V_0}{N}=-\frac{\sigma}{g_0^2}+\int\frac {d^3p_E}{(2\pi)^3}
            \ln[1+\frac{\sigma}{(p_\mu+A_\mu)(p_\mu+A_\mu)}].
\end{equation}
{}From this potential, it is clear that the constant $A_\mu$, which can be
gauged away, does not give rise to any physical result for the
renormalizable theory of the $CP^{N-1}$ model on $R^3$, and we will set
$A_\mu=0$. The effective potential of the $CP^{N-1}$ model can be treated
almost identically with that of the $O(N)$ model. The renormalization can be
done by demanding that
\begin{equation}
\frac{1}{N}\frac{\partial V_0}{\partial\sigma}\mid_{\sigma=M^2}
 =-\frac{1}{g^2},
\end{equation}
and then with the cutoff $\Lambda$ the effective potential is written as
\begin{eqnarray}
\frac{V_0}{N}&=&-\frac{\sigma}{g^2}
  -\sigma\int_{\mid p_E\mid<\Lambda}\frac{d^3p_E}{(2\pi)^3}\frac{1}{p_E^2+M^2}
 +\int_{\mid p_E\mid <\Lambda}\frac{d^3p_E}{(2\pi)^3}
                    \ln[1+\frac{\sigma}{p_E^2}]\\
    &=&\sigma[\frac{M}{4\pi}-\frac{1}{g^2}]-\frac{\sigma^{3/2}}{6\pi}
       +O(1/\Lambda).\nonumber
\end{eqnarray}
The phase structure from the effective potential is identical with that of
$O(N)$ model on $R^3$. The critical coupling constant is
\begin{equation}
g_c^2=\frac{4\pi}{M}.
\end{equation}
For $g>g_c$ the mass $m_0$ of $z$-particles are dynamically generated as
\begin{equation}
m_0=M-\frac{4\pi}{g^2},
\end{equation}
and from the saddle point analysis of effective action [2,5] one can find that
$<z>$ is zero which means that the model is in disordered phase. Recently it
has been suggested that the hedgehog-like instantons which exist in this
model may yield some interesting consequences for the properties of
antiferromagnet in disordered phase while there is no significant effect of
instanton in ordered phase [13].
For $g<g_c$, there is no dynamical mass generation and stationary point
exists with nonvanishing $<z>$ which means that the model is in ordered
phase.\newline

\noindent
{\bf B. The model on $R^2\times S^1$}\newline

The boundary condition of $z$-fields along the circulation of the circle $S^1$
again will be assumed as
\begin{equation}
z(x_1, x_2, x_3+L)= e^{i2\pi\delta} z(x_1, x_2, x_3)~~~(0\leq \delta <1).
\end{equation}
We can use the periodic $z$-field through the gauge  transformation
$z'=e^{-i2\pi\delta\frac{x_3}{L}}z$, however, with the changed gauge field
$A_\mu'=A_\mu+\frac{2\pi\delta}{L}\delta_{\mu,3}$. Instead of doing this, we
will assume a $\delta$ and gauge field $A_\mu=A_3\delta_{\mu,3}$.
The effective potential is then
\begin{equation}
\frac{V_L(\sigma,\delta)}{N}=-\frac{\sigma}{g_0^2}+\frac{1}{L}\sum_{n=-\infty}^\infty   \int\frac{d^2p_E}{(2\pi)^2}\ln[1+\frac{\sigma}
      {(\frac{2\pi}{L})^2(n+\delta+\frac{L}{2\pi}A_3)^2 +p_E^2}].
\end{equation}
Through the same methods for the $O(N)$ model, we obtain the renormalized
effective potential
\begin{equation}
V_L(\sigma,\delta)=V_0(\sigma)+NV_\Delta(\sigma,L,\delta),
\end{equation}
where
\begin{eqnarray}
V_\Delta &=&\frac{1}{\pi L^3}\sum_{n=1}^\infty\frac{\cos(2\pi\delta' n)}{n^3}
         (1-e^{-nL\sqrt{\sigma}})\nonumber\\
     & &-\frac{1}{\pi L^2}\sum_{n=1}^\infty \frac{\sqrt{\sigma}}{n^2}
          \cos(2\pi\delta' n) e^{-nL\sqrt{\sigma}}.
\end{eqnarray}
The $\delta'$ denotes $\delta+ LA_3/2\pi$ which may be justly termed
effective boundary condition parameter. $V_L$ depends only on $\delta'$ not
separately  on $\delta$ or $A_3$, which is a reflection of gauge invariance
on $R^2\times S^1$.

For the stationary point, $V_L$ must satisfy
\begin{eqnarray}
\frac{1}{N}\frac{\partial V_L}{\partial \sigma}\mid_{\sigma=m_L^2} &=&
      0\\
 &=&-\frac{1}{g^2}+\frac{1}{4\pi}[M-m_L]
   -\frac{1}{4\pi L}\ln[1-2\cos(2\pi\delta')e^{-Lm_L}+e^{-2Lm_L}]\nonumber
\end{eqnarray}
which is similar to the Eq.(15) of the $O(N)$ model. However, there is another
condition for the stationary point in $CP^{N-1}$ model:
\begin{eqnarray}
\frac{\partial V_L}{\partial A_3}&=&0\\
     &=& \frac{\partial \cos(2\pi\delta')}{\partial A_3}
      \frac{\partial V_L}{\partial \cos(2\pi\delta')}
     =-\frac{L}{2\pi}\sin(2\pi\delta')\frac{\partial V_L}{\partial
\cos(2\pi\delta')}
\nonumber\\
\end{eqnarray}
which is absent in the $O(N)$ model. The two conditions for stationary point
can be satisfied with
\begin{equation}
m_L=\frac{1}{L}\cosh^{-1}[1+\frac{\exp\{(M-\frac{4\pi}{g^2})L\}}{2}]
\end{equation}
and
\begin{equation}
\delta'=0~~~({\rm mod}~~~1).
\end{equation}

At the stationary point of the model on $R^2\times S^1$, $A_3$ is to be
arranged so that the effective boundary parameter is 0.  Therefore, the
dynamical mass generation takes place for any $\delta$ and $L$, which implies
the model is in disordered phase.

\begin{center}{\bf IV. CONCLUSION}\end{center}

We evaluate the large-$N$ effective potentials of the $O(N)$ model and the
$CP^{N-1}$ model. It is shown that the $O(N)$ model has rich phase structure
on $R^2\times S^1$ while the $CP^{N-1}$ model is always in disordered phase.

Recently we studied four-fermion interaction model on $R^2\times S^1$[7], where
we also obtained the rich phase structure similar to that of the $O(N)$
model. The phase structure of the $CP^{N-1}$ model is relatively very simple.
This is because the $CP^{N-1}$ model has Abelian gauge symmetry, and  the
presence
of Abelian gauge symmetry enforces the effective boundary condition
parameter to be 0 as in the Abelian gauge theory coupled to charged scalar
field. It suggests that on a nontrivial topology phase structure of a model
which has Abelian gauge field would be relatively simple. Through the similar
analysis of this paper it is easy to find that the $CP^{N-1}$ model on
$R^1\times S^1$ is in disordered phase.

The limit $\delta=0$ coincides with a case of finite temperature. In both
cases of the $O(N)$ model and the $CP^{N-1}$ model, we recover the well-known
result that, at any finite temperature, dynamical mass generations take place
and the models are in disordered phases  [5,9] ( see also
Ref.[11] ). For the $O(N)$ model, the $L_c$ ( $T_c$ ) of $g<g_c$ approaches to
$\infty$ ( 0 ) in this limit. In the previous analysis of the finite
temperature  $CP^{N-1}$ model [5] it was assumed that $A_3=0$, while our
results show that why it must be so. The $m_L$ in Eq.(35) is equal to the
finite temperature mass of Ref.[5] in the large-$N$ limit when
$L=\beta=1/k_B T$.

\vspace{1cm}
\begin{center}{\bf ACKNOWLEDGMENTS}\end{center}
The author thanks Professor Antal Jevicki, Professor Choonkyu Lee, Dr. J. Lee,
and Professor Kwang Sup Soh for very helpful discussions.
This research was supported in part by NON DIRECTED RESEARCH FUND, Korean
Research Foundation, 1993, and also by Korea Science and Engineering
Foundation.\newpage

\noindent
[1] I. Ya Aref'eva, Teor. Mat. Fiz 36,24 (1978); {\em ibid}. 36, 159 (1978);
      Ann. Phys. (N.Y.) 117, 39 (1979).\newline
[2] I. Ya Aref'eva and S. I. Azakov, Nucl. Phys. B162, 298 (1980).\newline
[3] B. Rosenstein, B.J. Warr, and S.H. Park, Phys. Rev. Lett. 62, 1433 (1989).
   \newline
[4] E. Brezin and J. Zinn-Justin, Phys. Rev. Lett. 36, 691 (1976); Phys. Rev.
    B14 3110 (1976); F.D.M. Haldane, {\em ibid.} 50, 1153 (1983); W. G. Wen and
    A. Zee, Phys. Rev. Lett. 61, 1025 (1988); F.D.M. Haldane, {\em ibid.} 61,
1209
    (1988); For review, see E. Fradkin, {\em Field Theories of Condensed
    Matter System} (Addison-Wesley, Redwood, 1991) and references therein.
\newline
[5] I. Ichinose and H. Yamamoto, Mod. Phys. Lett. 5, 1373 (1990).\newline
[6] A. Higuchi and L. Parker, Phys. Rev. D37, 2853 (1988).  \newline
[7] D.Y. Song, Phys. Rev. D48, 3925 (1993).     \newline
[8] A. D'Adda, M. L\"{u}scher, and P.Di Vecchia, Nucl. Phys. B146, 63 (1978);
    E. Witten, {\em ibid}. B149, 285 (1979).\newline
[9] B. Rosenstein, B.J. Warr, and S.H. Park, Nucl. Phys. B336, 435 (1990).
                     \newline
[10] S. Coleman, {\em Aspects of Symmetry} (Cambridge University Press, London,
    1985).\newline
[11] The effective potential we construct is quite different from that in
     Ref.[2] where the potential is in terms of $<\psi\mid n\mid\psi>$.
     In Ref.[9], the effective potential has been given in terms of
     magnetization and used in studying the model at finite temperature,
     which are also quite different from the potential in text. It is
     not clear to us how to understand the relations of these
     different potentials.\newline
[12] R. Jackiw, Phys. Rev. D9, 1686 (1974).\newline
[13] G. Murthy, Phys. Rev. Lett. 67, 911(1991); G. Murthy and S. Sachdev, Nucl.
Phys. B344, 557 (1990).

\end{document}